\documentclass[aps,preprint,superscriptaddress]{revtex4-2}
               
\usepackage{graphicx}   
\usepackage{dcolumn}    
\usepackage{bm}         
\usepackage{amsmath}
\usepackage{amssymb}
\usepackage[usenames]{color}      

\begin{document}

\title{Realizing the ultimate scaling of the convection turbulence by spatially decoupling the thermal and viscous boundary layers}

\author{Shufan Zou}
\affiliation{State Key Laboratory for Turbulence and Complex Systems, Department of Mechanics and Engineering Science, College of Engineering, Peking University, Beijing 100871, China}

\author{Yantao Yang}\email{yantao.yang@pku.edu.cn}
\affiliation{State Key Laboratory for Turbulence and Complex Systems, Department of Mechanics and Engineering Science, College of Engineering, Peking University, Beijing 100871, China}
\affiliation{Beijing Innovation Center for Engineering Science and Advanced Technology, Peking University, Beijing 100871, China}
\affiliation{Institute of Ocean Research, Peking University, Beijing 100871, China}
       
\date{\today}
	
\begin{abstract}

Turbulent convection plays a crucial role in many natural environments, ranging from Earth ocean, mantle and outer core, to various astrophysical systems. For such flows with extremely strong thermal driving, an ultimate scaling was proposed for the heat flux and velocity. Despite numerous experimental and numerical studies, a conclusive observation of the ultimate regime has not been reached yet. Here we show that the ultimate scaling can be perfectly realized once the thermal boundary layer is fully decoupled from the viscous boundary layer and locates inside the turbulent bulk. The heat flux can be greatly enhanced by one order of magnitude. Our results provide concrete evidences for the appearance of the ultimate state when the entire thermal boundary layer is embedded in the turbulent region, which is probably the case in many natural convection systems.

\end{abstract}

\maketitle

\bigskip

\section{Introduction}

The paradigmatic model for turbulent convection is the Rayleigh-B\'{e}nard (RB) flow, namely the buoyancy-driven flow within a fluid layer heated from below and cooled from above. Such flow system can be found in many natural environments~\cite{spiegel1971,tackley1993,marshall1999,heimpel2005}. For a given fluid with certain Prandtl number $Pr=\nu/\kappa$, the driving force is measured by the Rayleigh number $Ra=g \alpha \Delta H^3 / \nu\kappa$. Here, $\nu$ is kinematic viscosity, $\kappa$ is thermal diffusivity, $g$ is the gravitational acceleration, $\alpha$ is thermal expansion coefficient, $\Delta$ is the temperature difference across the layer, and $H$ is the layer height, respectively. The most fundamental question is how the heat flux and the turbulent level of the flow depend on $Ra$ and $Pr$~\cite{ahlers2009,chilla2012,xia2013,plumley2019}. Here, the heat flux is usually measured by the Nusselt number $Nu$, i.e. the ratio between the convective flux to the conductive flux. The turbulent level of the flow by the Reynolds number $Re=U H / \nu$, with $U$ being some characteristic value of the flow velocity.

In RB flow, thermal and viscous boundary layers develop next to the top and bottom plates, with the convective bulk in between and full of thermal plumes. In most early experiments and simulations the viscous boundary layer is laminar, and the heat flux follows the ``classic'' scaling law $Nu \sim Ra^\gamma$ with $\gamma\approx1/3$~\cite{malkus1954}. It was hypothesized that when $Ra$ is large enough, saying the thermal driving is extremely strong, the viscous boundary layer becomes fully turbulent due to the shearing exposed by the vigorous bulk flow, and the system enters the ``ultimate'' state where the heat flux no longer depends on the viscosity of the fluid~\cite{kraichnan1962,gl2000,gl2001}. The heat flux then follows the ultimate scaling with $\gamma=1/2$, which predicts $Nu$ many orders of magnitude larger than the classic scaling at very large $Ra$ as in various natural environments.

Tremendous efforts have been made to achieve the ultimate scaling in RB flow by both experiments and simulations. Fluid with very small $Pr$, such as mercury, is used so that the thermal boundary layer is much thicker than the viscous one and extends into the turbulent bulk~\cite{chavanne1997,glazier1999,niemela2000}. Recent state-of-art simulations and experiments push $Ra$ to very high values~\cite{niemela2000,he2012,zhu2018prl,iyer2020}. Some of these studies reported evidences of the transition to the ultimate regime~\cite{chavanne1997,he2012,zhu2018prl}. Wall roughness was introduced to trigger the transition of the momentum boundary layer to fully turbulent state, but this only works for a limit range of $Ra$ for a given wall roughness~\cite{zhu2017,zhu2019}. Homogeneous convection in fully periodic domain exhibits the ultimate scaling, since in such flow all boundary layers are removed and only the turbulent bulk plays a role~\cite{lohse2003,calzavarini2005}. The radiative convection with internal heating and the convection with background oscillation can both greatly enhance the exponent $\gamma$ from the classic-scaling value, but such treatments change the dynamics of the convection system by introducing extra source terms for the thermal field or the momentum field~\cite{lepot2018,bouillaut2019,creyssels2020,wang2020}.

\section{Governing equations and numerical methods}

Here by specially designed numerical experiments, we confirm the physical conjecture of the ultimate paradigm. Once the whole thermal boundary layer is spatially decoupled from the viscous one and locate entirely in the turbulent flow region, the ultimate scaling of the heat flux can be perfectly realized even at low to moderate Rayleigh numbers. Specifically, we consider the RB flow between two parallel plates which are perpendicular to the direction of the gravity and separated by a distance $H$. A homogeneous layer of height $h$ is introduced at top and bottom of the domain. Within this layer the temperature is uniform and equal to the temperature of the adjacent plate, which can be readily realized by using an immersed-boundary technique. The buoyancy-force term takes effect only between the two homogeneous layers. By doing so the thermal boundary layer is lifted by a height $h$ from the plate where the viscous boundary layer occurs. If $h$ is large enough, the two boundary layers can be fully decoupled.

Specifically, we consider a fluid layer bounded by two parallel plates which are perpendicular to the gravity and separated by a distance $H$. The Oberbeck-Bbuossinesq approximation is utilized to take care of the buoyancy effects. First, the fluid density depends linearly on the scalar field as $\rho = \rho_0(1-\alpha \theta)$. Here $\rho_0$ is the reference density at the reference temperature $T_0$, and $\theta$ is the temperature deviation from $T_0$. Second, the variation of density is relatively small so that all the fluid properties can be treated as constants, and only the buoyancy force needs to be included. The governing equations then read
\begin{eqnarray}
  \partial_t u_i + \mathbf{u}\cdot \nabla \mathbf{u} &=& 
       - \frac{1}{\rho} \nabla p + \nu\,\nabla^2 \mathbf{u} 
       + \alpha\,\theta\,\mathbf{g}, \label{eq:momen}  \\
  \partial_t \theta + \mathbf{u}\cdot\nabla \theta &=& 
       \kappa\,\nabla^2 \theta,   \label{eq:temper}  \\
  \nabla\cdot\mathbf{u} &=& 0. \label{eq:continu}
\end{eqnarray}
Here $\mathbf{u}$ is three dimensional velocity, $p$ is pressure, and $\partial_t$ is the partial derivative with respect to time, respectively. The above equations are non-dimensionalized by the layer height $H$, the temperature difference between two plates $\Delta$, and the free-fall velocity $U=\sqrt{g \alpha \Delta H}$. The two plates are no-slip and the periodic boundary conditions are applied to the two horizontal directions.

Direct numerical simulations were conducted with the in-house code, which employs the finite difference scheme and a fraction-time-step method. The code has been tested for various wall-turbulence and convection turbulence~\cite{multigrid2015}. To introduce the homogeneous thermal layer, an extra source term is included in the temperature equation~\eqref{eq:temper} based on the idea of the immersed boundary method~\cite{fadlun2000}. Also the buoyancy force in the momentum equation~\eqref{eq:momen} is turned off inside the homogeneous thermal layer. Grid-independence test has been conducted in order to ensure the resolution is adequate. The three-dimensional volume rendering in figure~1 is generated by the open source software VisIt~\cite{visit}.

\section{Main results}

\begin{figure}
	\centering
	\includegraphics[width=\textwidth]{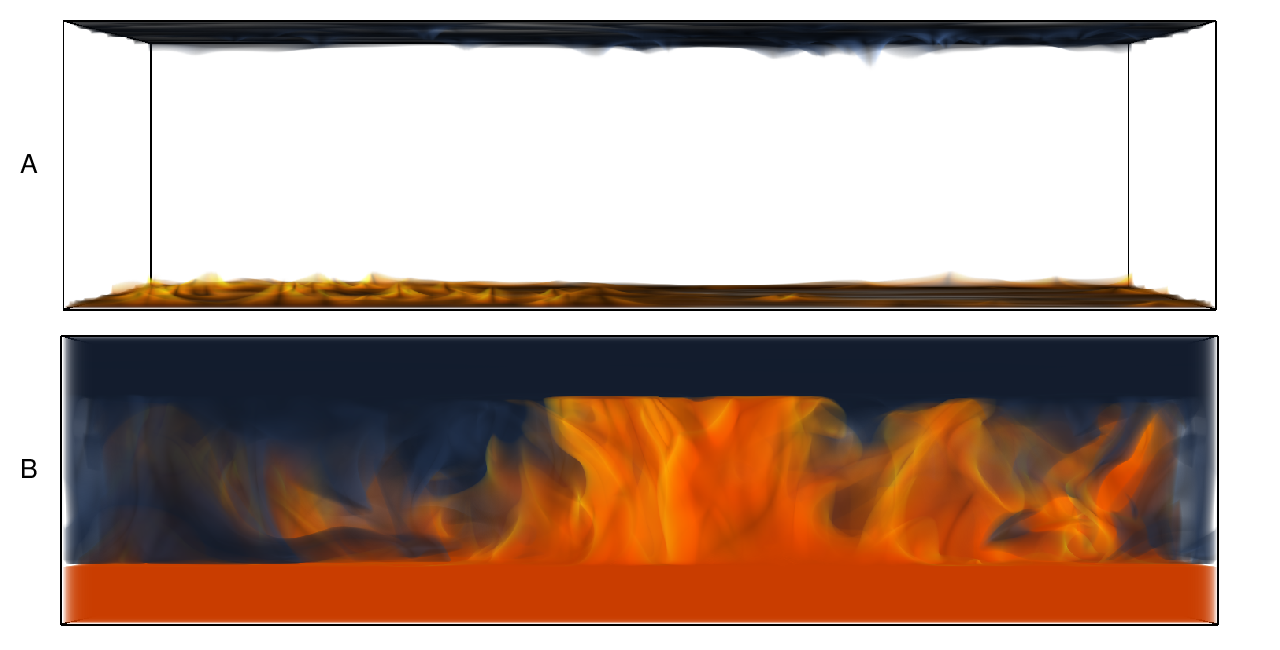}%
	\caption{Volume rendering of the thermal field at $Ra=10^6$ and $Pr=1$. Yellow and blue indicate the temperature larger than $90\%\Delta$ and smaller than $10\%\Delta$, respectively. A: the normal Rayleigh-B\'{e}nard flow. B: the modified flow with two homogeneous thermal layers of the height $h=0.2H$. The flow in B has much stronger thermal plumes in the bulk, while for the flow in A the large temperature anomaly can hardly be transported into the bulk.} 
	\label{fig:3d}
\end{figure}

In figure~\ref{fig:3d} we compare one thermal field of the normal RB flow and that with homogeneous layers at $Ra=10^6$ and $Pr=1$. Here $h$ is chosen as $0.2H$. We only show the fluid with temperature larger than $90\%\Delta$ and smaller than $10\%\Delta$. For normal RB flow at this relatively low $Ra$, the large thermal anomaly cannot be transported very far from the plates, which limits the global heat flux. On the contrary, with two homogeneous thermal layers the thermal plumes with large anomaly are significantly stronger, and spread over the entire convection bulk. Accordingly, the heat flux is greatly enhanced in the flow with homogeneous thermal layers.

\begin{figure}
	\centering
	\includegraphics[width=\textwidth]{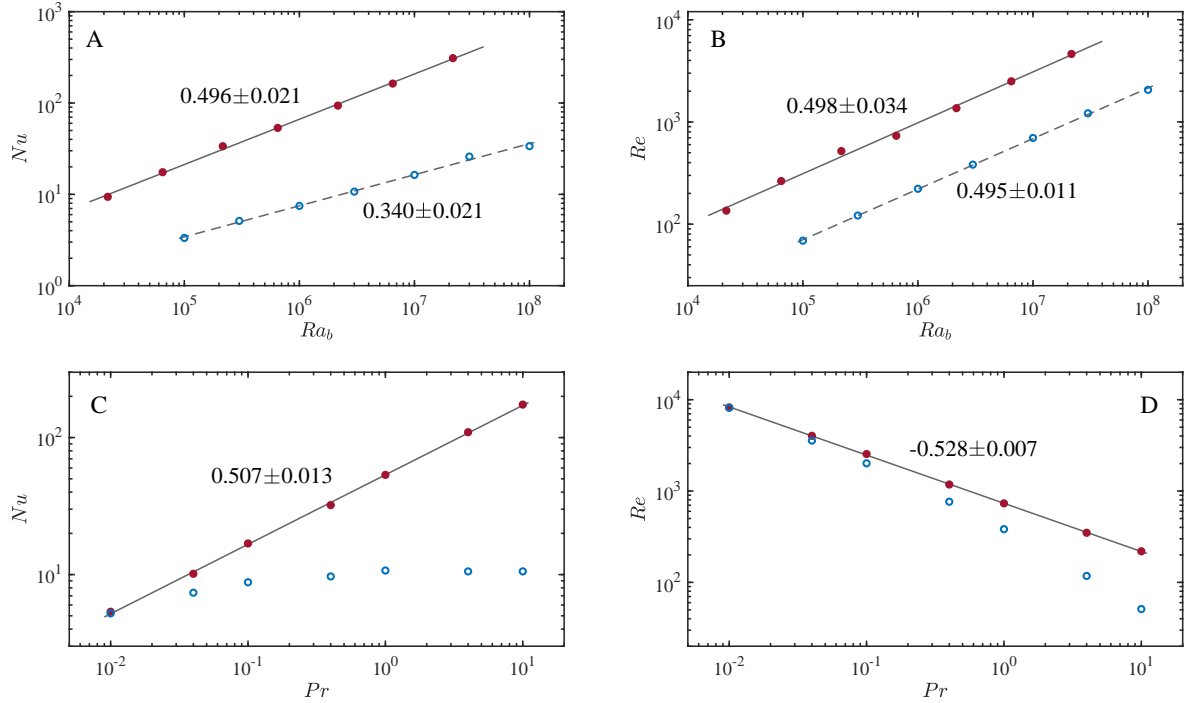}%
	\caption{The ultimate scaling of the heat flux and flow velocity. Solid symbols and lines depict the ultimate scalings of the modified flow with homogeneous thermal layers. Open symbols and dashed lines show the normal RB flow and the classic scaling. The corresponding exponents are determined by linear fitting. A and B: The Nusselt number $Nu$ and Reynolds number $Re$ versus the increasing bulk Rayleigh number $Ra_b$ for a fixed Prandtl number $Pr=1$. C and D: $Nu$ and $Re$ versus the increasing $Pr$ for fixed $Ra=3\times10^6$.} 
	\label{fig:scaling}
\end{figure}

We then present the main results of the current study, i.e. the ultimate scalings of heat flux and flow velocity, which in their full form read
\begin{equation}\label{eq:scaling}
  Nu\sim Pr^{1/2}Ra_b^{1/2},\quad Re\sim Pr^{-1/2}Ra_b^{1/2}.
\end{equation}
To account for the height of two homogeneous thermal layers, the bulk Rayleigh number is calculated as $Ra_b = Ra (1-2h/H)^3$. The Nusselt number is defined as $Nu=q(H-2h)/(\kappa\Delta)$ with $q$ being the total dimensional heat flux. When $h=0$, both definitions recover the usual form of RB flow. The Reynolds number is $Re=u_{rms}H/\nu$ and $u_{rms}$ is the root-mean-square velocity of the whole domain, since the flow motion is confined by two plates, including the homogeneous thermal layers. We first look at the dependences of $Nu$ and $Re$ on $Ra_b$ for fixed $Pr=1$. Three decades of $Ra$ are covered from $10^5$ up to $10^8$. The height of the homogeneous thermal layer is set at $h=0.2H$ so that for all simulations the momentum boundary layer is fully embedded inside the homogeneous thermal layer. Figures~\ref{fig:scaling}A and \ref{fig:scaling}B show that both $Nu$ and $Re$ follow the ultimate scaling law \eqref{eq:scaling} perfectly, even though the Rayleigh number is not very large. Moreover, both the heat flux and flow velocity are greatly enhanced by introducing the homogeneous thermal layers. For $Ra_b\ge10^6$, the modified flow generates a heat flux more than one order of magnitude higher than that for the normal RB flow. For instance, the Nusselt number is comparable for the modified flow at $Ra\approx10^8$ and the normal RB flow at $Ra\approx10^{11}$. 

For the second group of simulations, we fix $Ra=3\times10^6$ and increase $Pr$ from $0.01$ to $10$, again covering three orders of magnitude. The height of the homogeneous layers is $h=0.2H$. As shown in figures~\ref{fig:scaling}C and \ref{fig:scaling}D, the behaviors of $Nu$ and $Re$ are also very close to the ultimate scaling~\eqref{eq:scaling}. The scaling exponent for $Nu$ given by the linear fitting is indistinguishable from the theoretically predicted value $1/2$. The exponent for $Re$ obtained from the numerical data is slightly different from $-1/2$. For the moderate $Ra=3\times10^6$ considered here, the flow is not fully turbulent at high Prandtl number regime. Therefore $Re$ decreases faster than that predicted by the ultimate scaling. Another interesting observation is that at low $Pr$ region, the difference between the modified flow and the normal RB flow is very small. This is expectable since for small $Pr$, the viscous boundary layer is thinner than the thermal one in the normal RB flow, and part of the thermal boundary layer is already in the turbulent bulk.  

\begin{figure} 
	\centering
	\includegraphics[width=\textwidth]{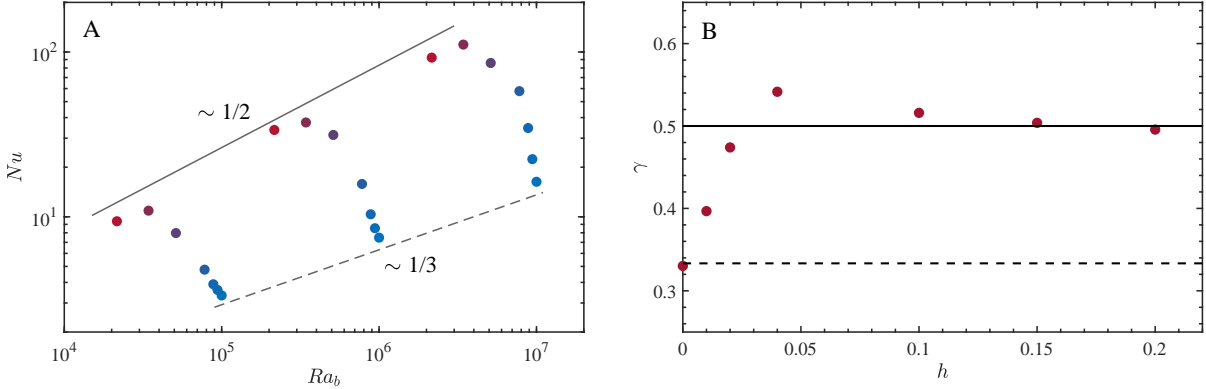}%
	\caption{The effect of the thermal layer height on the heat-flux scaling. A: The heat flux versus $Ra_b$ for different $h$, which increases from $0$ to $0.2$ as color changes from blue to red. B: The scaling exponent $\gamma$ versus the height $h$. In both panels the solid and dashed lines mark the ultimate scaling with the exponent $1/2$ and the classic scaling with the exponent $1/3$, respectively.} 
	\label{fig:h}
\end{figure}
To further demonstrate that the ultimate scaling is satisfied when the thermal and viscous boundary layers are decoupled from each other, we run simulations with increasing $h$ for three different Rayleigh numbers $Ra=10^5$, $10^6$, and $10^7$. The Prandtl number is fixed at $Pr=1$. We focus on the behaviors of Nusselt number, which are shown in figure~\ref{fig:h}. As $h$ increases, i.e. the thermal boundary layers are gradually decoupled from the viscous ones, the heat flux is enhanced and the scaling exponent transits from the classic value of $1/3$ to the ultimate value $1/2$. In figure~\ref{fig:h} we plot the scaling exponent $\gamma$ versus the height of the homogeneous thermal layer $h$. For small $h$, $\gamma$ increases rapidly to a value larger than $1/2$. At this overshoot the thermal boundary layer already enters the turbulent bulk for large $Ra$ but is still coupled with the viscous boundary layer for low $Ra$. When $h$ is large enough, for all three $Ra$'s the thermal boundary layer is fully within the turbulent region, and the exponent $\gamma$ gradually approaches the ultimate value $1/2$.

\section {Conclusion and discussion}

In summary, we demonstrate that once the thermal boundary layers decouple with the viscous ones and locate within the turbulent convection bulk, the ultimate scaling of the heat flux and flow velocity can be perfectly realized even at relatively low Rayleigh numbers. Our results support the physical picture of the ultimate state of convection turbulence, namely when the momentum boundary layers become fully turbulent, the heat flux is independent of the fluid viscosity. In line with the unifying model for RB convection~\cite{gl2000,gl2001}, the current flow corresponds to the IV$_l$ regime. Compared to the homogeneous convection~\cite{lohse2003,calzavarini2005}, the current flow configuration is more close to the ultimate regime since the thermal boundary layers are included and the ultimate scaling is achieved.

A definitive proof for the existence of the ultimate regime relies on the direct observation of the ultimate scaling in the experiments and simulations of the normal RB convection. Nevertheless, one can anticipate that, based on the present study, convection flows will eventually enter the ultimate state when the boundary layer region becomes fully turbulent at extremely high thermal driving. Such conditions are very likely satisfied in natural systems such as the Earth outer core and stars. Transition towards the ultimate regime has been reported in recent studies~\cite{he2012,zhu2018prl}, and we may expect that a fully developed ultimate state will be observed in experiments in near future.

\

{\it Acknowledgements:} This work is supported by the Major Research Plan of National Natural Science Foundation of China for Turbulent Structures under the Grants 91852107 and 91752202. Y.Yang also acknowledges the partial support from the Strategic Priority Research Program of Chinese Academy of Sciences under the Grant No. XDB42000000.

\end{document}